\documentstyle[prl,aps,multicol]{revtex}

\renewcommand{\narrowtext}{\begin{multicols}{2} \global\columnwidth20.5pc}
\renewcommand{\widetext}{\end{multicols} \global\columnwidth42.5pc}
\begin{document}
\draft
\title{Novel universal correlations in invariant random-matrix models}
\author{E. Kanzieper and V. Freilikher}
\address{The Jack and Pearl Resnick Institute of Advanced Technology,\\
Department of Physics, Bar-Ilan University, Ramat-Gan 52900, Israel}
\date{December 15, 1996}
\maketitle

\begin{abstract}
We show that eigenvalue correlations in unitary-invariant ensembles of large
random matrices adhere to novel universal laws that only depend on a
multicriticality of the bulk density of states near the soft edge of the
spectrum. Our consideration is based on the previously unknown observation
that genuine density of states and $n-$point correlation function are
completely determined by the Dyson's density analytically continued onto the
whole real axis.
\end{abstract}

\pacs{\tt chao-dyn/9701006}

\narrowtext
Random matrices have been introduced in a physical context since the works
by Wigner \cite{Wigner} and Dyson \cite{Dyson}. Initially proposed as an
effective phenomenological model for description of the higher excitations
in nuclei \cite{Nuclei}, the invariant ensembles of large random matrices
found numerous applications in very diverse fields of physics such as
two-dimensional quantum gravity \cite{Migdal}, quantum chaos \cite{Chaos},
and mesoscopic physics \cite{Mesoscopics}. Apparently, this ubiquity owes
its origin to the very idea of the construction of the invariant one-matrix
model \cite{Mehta}, which only reflects the fundamental symmetry
(orthogonal, unitary or symplectic) of underlying physical system/phenomenon
but discards its (irrelevant) microscopic details. Since the symmetry
constraints follow from the first principles, even a rather crude matrix
model allows identification of universal features which persist for a
variety of systems with the same symmetry. This circumstance emphasizes the
importance of the study of universality intrinsic in random matrices.

The simplest invariant random-matrix model is defined by the probability
density 
\begin{equation}
P\left[ {\bf H}\right] =\frac 1{{\cal Z}_N}\exp \left\{ -%
\mathop{\rm Tr}
V\left[ {\bf H}\right] \right\}  \label{eq.01}
\end{equation}
of the entries $H_{ij}$ of the $N\times N$ random matrix ${\bf H}$, where
the function $V\left[ {\bf H}\right] $ referred to as ``confinement
potential'' must ensure existence of the partition function ${\cal Z}_N$, $%
N\gg 1$.

In the following we restrict our consideration to the unitary invariant, $%
U\left( N\right) $, matrix model. Nowadays it is widely believed that $%
U\left( N\right) $ invariant ensembles of large random matrices with rather
strong level confinement may exhibit {\it three} different types of locally
universal eigenlevel correlations which are characterized by the
appropriately scaled two-point kernels.

$\bullet $ {\it Bulk scaling limit} is associated with a spectrum range
where the confinement potential is well-behaved, and density of levels can
approximately be taken as a constant. It has been proven in Refs. \cite
{Brezin-Zee,Weidenmueller,Kanzieper} that for rather strong confinement
potentials \cite{remark-1} the two-point kernel follows the {\it universal
sine law} 
\begin{equation}
K_{\text{bulk}}\left( s,s^{\prime }\right) =\frac{\sin \left[ \pi \left(
s-s^{\prime }\right) \right] }{\pi \left( s-s^{\prime }\right) }.
\label{eq.02}
\end{equation}
Here scaling variable $s$ is measured in the units of the mean level
spacing: $s=\varepsilon /\Delta _N$.

$\bullet $ {\it Origin scaling limit }deals with that part of the spectrum
where confinement potential displays logarithmic singularity: $V\left(
\varepsilon \right) \rightarrow V\left( \varepsilon \right) -\alpha \log
\left| \varepsilon \right| $. In the vicinity of the singularity $%
\varepsilon =0$ the two-point kernel takes the {\it universal Bessel law} 
\cite{Nishigaki}: 
\[
K_{\text{orig}}\left( s,s^{\prime }\right) 
\]
\begin{equation}
=\frac \pi 2\sqrt{ss^{\prime }}\frac{J_{\alpha +\frac 12}\left( \pi s\right)
J_{\alpha -\frac 12}\left( \pi s^{\prime }\right) -J_{\alpha -\frac 12%
}\left( \pi s\right) J_{\alpha +\frac 12}\left( \pi s^{\prime }\right) }{%
s-s^{\prime }}.  \label{eq.03}
\end{equation}
Here $s$ is scaled by the level spacing near the origin, $s=\varepsilon
/\Delta _N\left( 0\right) $.

$\bullet $ {\it Soft-edge scaling limit}, relevant to the tail of eigenvalue
support where crossover occurs from a non-zero density of states to a
vanishing one \cite{Bowick}, has been only investigated for Gaussian unitary
ensemble (GUE) \cite{Forrester}, and quite recently for $U\left( N\right) $
invariant ensembles of large random matrices associated with quartic and
sextic confinement potentials \cite{Airy}. It has been found that in the
soft-edge scaling limit in all these ensembles the two-point kernels follow 
{\it Airy law} 
\begin{equation}
K_{\text{soft}}\left( s,s^{\prime }\right) =\frac{%
\mathop{\rm Ai}
\left( s\right) 
\mathop{\rm Ai}
^{\prime }\left( s^{\prime }\right) -%
\mathop{\rm Ai}
\left( s^{\prime }\right) 
\mathop{\rm Ai}
^{\prime }\left( s\right) }{s-s^{\prime }}.  \label{eq.04}
\end{equation}
Here $s\propto N^{2/3}\cdot \left( \varepsilon /D_N-1\right) $ with $D_N$
being the endpoint of the spectrum.

Whereas universality in the spectrum bulk and near its origin has rigorously
been proven for a wide class of strong symmetric confinement potentials, the
supposed universality of the Airy kernel has not been proven.

Our aim here is to demonstrate that the Airy correlations, Eq. (\ref{eq.04}%
), being universal for a wide class of matrix models Eq. (\ref{eq.01}), are
indeed a particular case of more general novel universal correlations which
are represented by the scaled $m-$th multicritical two-point kernel%
\widetext
\begin{equation}
K_{\text{soft}}^{\left( m\right) }\left( s,s^{\prime }\right) =\frac{%
\mathop{\rm G}
\left( s|\nu ^{*}\right) 
\mathop{\rm G}
^{\prime }\left( s^{\prime }|\nu ^{*}\right) \cdot s^{\frac 32-\nu ^{*}}-%
\mathop{\rm G}
\left( s^{\prime }|\nu ^{*}\right) 
\mathop{\rm G}
^{\prime }\left( s|\nu ^{*}\right) \cdot \left( s^{\prime }\right) ^{\frac 32%
-\nu ^{*}}}{s-s^{\prime }}\text{,}  \label{eq.05}
\end{equation}
where the function $%
\mathop{\rm G}
$ is expressed through the Bessel functions as 
\begin{eqnarray}
\mathop{\rm G}
\left( s|\nu ^{*}\right)  &=&\frac 1{2\sqrt{\nu ^{*}}}\left[ \sin \left( 
\frac \pi {4\nu ^{*}}\right) +\left( -1\right) ^{\nu ^{*}-\frac 32}\right]
^{-1/2}  \label{eq.06} \\
&&\ \times \left\{ 
\begin{array}{ll}
s^{\frac 12\left( \nu ^{*}-\frac 12\right) }\left[ I_{-\frac 12\left( 1-%
\frac 1{2\nu ^{*}}\right) }\left( \frac{s^{\nu ^{*}}}{\nu ^{*}}\right) -I_{%
\frac 12\left( 1-\frac 1{2\nu ^{*}}\right) }\left( \frac{s^{\nu ^{*}}}{\nu
^{*}}\right) \right] , & s>0, \\ 
\left| s\right| ^{\frac 12\left( \nu ^{*}-\frac 12\right) }\left[ J_{-\frac 1%
2\left( 1-\frac 1{2\nu ^{*}}\right) }\left( \frac{\left| s\right| ^{\nu ^{*}}%
}{\nu ^{*}}\right) +\left( -1\right) ^{\nu ^{*}-\frac 32}J_{\frac 12\left( 1-%
\frac 1{2\nu ^{*}}\right) }\left( \frac{\left| s\right| ^{\nu ^{*}}}{\nu ^{*}%
}\right) \right] , & s<0,
\end{array}
\right.   \nonumber
\end{eqnarray}
\narrowtext\noindent
and parameter $\nu ^{*}$ is determined by the even critical index \cite
{Bowick} $m=0,2,4,$ etc. of the matrix model: 
\begin{equation}
\nu ^{*}=m+\frac 32.  \label{eq.07}
\end{equation}
Note that the critical index $m$ is completely determined by the type of
singularity of the density of states near the soft edge \cite{Bowick}: $\nu
_N\left( \varepsilon \right) =\left\langle 
\mathop{\rm Tr}
\delta \left( \varepsilon -{\bf H}\right) \right\rangle \propto \left(
1-\varepsilon ^2/D_N^2\right) ^{m+1/2}$.

Equations (\ref{eq.05}) - (\ref{eq.07}) together with Eqs. (\ref{eq.25}) and
(\ref{eq.26}) below are the main results of the paper. Although we
concentrate our attention on the problem of eigenvalue correlations near the
soft edge, the treatment we present here is quite general being relevant to
arbitrary spectrum range.

{\bf A. }Within the orthogonal polynomial technique the two-point kernel $%
K_N\left( \varepsilon ,\varepsilon ^{\prime }\right) $ determining the $n-$%
point correlation function $R_n$ for eigenvalue spectrum of large random
matrices, $R_n\left( \varepsilon _1,...,\varepsilon _n\right) =\det \left[
K_N\left( \varepsilon _i,\varepsilon _j\right) \right] _{i,j=1...n}$, can be
written through the fictitious ``wavefunctions'' $\psi _n\left( \varepsilon
\right) $ as 
\begin{equation}
K_N\left( \varepsilon ,\varepsilon ^{\prime }\right) =c_N\frac{\psi _N\left(
\varepsilon ^{\prime }\right) \psi _{N-1}\left( \varepsilon \right) -\psi
_N\left( \varepsilon \right) \psi _{N-1}\left( \varepsilon ^{\prime }\right) 
}{\varepsilon ^{\prime }-\varepsilon }\text{.}  \label{eq.08}
\end{equation}
Here $c_N$ is the recurrence coefficient entering three-term recurrence
equation 
\begin{equation}
\begin{array}{lll}
\varepsilon P_{n-1}=c_nP_n+c_{n-1}P_{n-2}, & P_0\left( \varepsilon \right)
=1, & P_1\left( \varepsilon \right) =\varepsilon ,
\end{array}
\label{eq.09}
\end{equation}
for polynomials $P_n$ orthogonal on the whole real axis {\bf R}: 
\begin{equation}
\int d\alpha \left( \varepsilon \right) P_n\left( \varepsilon \right)
P_m\left( \varepsilon \right) =\delta _{nm},  \label{eq.10}
\end{equation}
and the ``wavefunction'' $\psi _n\left( \varepsilon \right) =P_n\left(
\varepsilon \right) \exp \left\{ -V\left( \varepsilon \right) \right\} $.
The measure $d\alpha \left( \varepsilon \right) $ $=\exp \left\{ -2V\left(
\varepsilon \right) \right\} d\varepsilon $ is completely determined by
symmetric confinement potential 
\begin{equation}
V\left( \varepsilon \right) =\sum_{k=1}^p\frac{d_k}{2k}\varepsilon ^{2k}
\label{eq.11}
\end{equation}
with $d_p>0$. The signs of the rest $d_k$'s can be arbitrary but they should
lead to an eigenvalue density supported on a single connected interval $%
\left( -D_N,+D_N\right) $.

To study the eigenvalue correlations in the random matrix ensemble with
confinement potential Eq. (\ref{eq.11}) we note that a three-term recurrence
equation for orthogonal polynomials $P_n\left( \varepsilon \right) $ can be
mapped onto a second order differential equation for these orthogonal
polynomials and/or corresponding wavefunctions $\psi _n\left( \varepsilon
\right) $. This was already observed for the first time by J. Shohat in 1930 
\cite{Shohat}. Considerably later Shohat's idea was developed by Bonan and
Clark \cite{Bonan-Clark}. The simple and elegant method proposed in Refs. 
\cite{Shohat,Bonan-Clark} turns out to be a very general and powerful one
for analysis of spectral properties possessed by large random matrices.

To map Eq. (\ref{eq.09}) onto a second order differential equation for $\psi
_n$, we note that the following identity takes place: 
\begin{equation}
\frac{dP_n}{d\varepsilon }=A_n\left( \varepsilon \right) P_{n-1}-B_n\left(
\varepsilon \right) P_n\text{,}  \label{eq.12}
\end{equation}
where the functions $A_n\left( \varepsilon \right) $ and $B_n\left(
\varepsilon \right) $ can be found from consideration below. Since $%
dP_n/d\varepsilon $ is a polynomial of the degree $n-1$ it can be
represented \cite{Szego} through the Fourier expansion in terms of the
kernel $K_n\left( t,\varepsilon \right) =\sum_{k=0}^{n-1}P_k\left( t\right)
P_k\left( \varepsilon \right) $ as follows: 
\begin{equation}
\frac{dP_n}{d\varepsilon }=\int d\alpha \left( t\right) \frac{dP_n}{dt}%
K_n\left( t,\varepsilon \right) \text{.}  \label{eq.13}
\end{equation}
Integrating by part we obtain that 
\begin{equation}
\frac{dP_n}{d\varepsilon }=2\int d\alpha \left( t\right) K_n\left(
t,\varepsilon \right) \left( \frac{dV}{dt}-\frac{dV}{d\varepsilon }\right)
P_n\left( t\right) \text{.}  \label{eq.14}
\end{equation}
Now, making use of the Christoffel-Darboux theorem \cite{Szego} we conclude
that unknown functions $A_n$ and $B_n$ in Eq. (\ref{eq.12}) are 
\begin{equation}
A_n\left( \varepsilon \right) =2c_n\int d\alpha \left( t\right) \frac{%
V^{\prime }\left( t\right) -V^{\prime }\left( \varepsilon \right) }{%
t-\varepsilon }P_n^2\left( t\right) \text{,}  \label{eq.15}
\end{equation}
\begin{equation}
B_n\left( \varepsilon \right) =2c_n\int d\alpha \left( t\right) \frac{%
V^{\prime }\left( t\right) -V^{\prime }\left( \varepsilon \right) }{%
t-\varepsilon }P_n\left( t\right) P_{n-1}\left( t\right) .  \label{eq.16}
\end{equation}

At first glance representations Eqs. (\ref{eq.15}) and (\ref{eq.16}) are
rather useless as far as they involve the same orthogonal polynomials which
enter Eq. (\ref{eq.12}). Nevertheless these expressions do allow us to get
the functions $A_n$ and $B_n$ in closed forms directly related to
confinement potential and to endpoint of eigenvalue spectrum. Restricting
our following consideration to large indices $n=N\gg 1$ we reduce Eq. (\ref
{eq.09}) to the asymptotic form 
\begin{equation}
\varepsilon P_N=c_N\left( P_{N+1}+P_{N-1}\right) \text{,}  \label{eq.17}
\end{equation}
where 
\begin{equation}
\varepsilon ^\lambda P_N=\sum_{j=0}^\lambda \left( 
\begin{array}{l}
\lambda  \\ 
j
\end{array}
\right) c_N^\lambda P_{N+2j-\lambda }\text{, }\lambda \geq 0\text{.}
\label{eq.18}
\end{equation}
Substituting $V\left( \varepsilon \right) $ given by Eq. (\ref{eq.11}) into
Eq. (\ref{eq.15}) yields 
\begin{equation}
A_N\left( \varepsilon \right) =2c_N\sum_{k=1}^p\sum_{\lambda
=1}^{2k-1}d_k\varepsilon ^{\lambda -1}\int d\alpha \left( t\right)
P_N^2\left( t\right) t^{2k-\lambda -1}\text{.}  \label{eq.19}
\end{equation}
Then, taking into account Eq. (\ref{eq.18}) as well as the orthogonality of $%
P_n$ we arrive at the expression defined for {\it arbitrary} $\varepsilon $: 
\begin{equation}
A_N\left( \varepsilon \right) =2c_N\sum_{k=1}^pd_k\sum_{\lambda =1}^k\left( 
\begin{array}{l}
2\left( k-\lambda \right)  \\ 
k-\lambda 
\end{array}
\right) c_N^{2k-\lambda }\varepsilon ^{2\lambda -2}.  \label{eq.20}
\end{equation}

It is easy to verify that $A_N\left( \varepsilon \right) $ can be
represented through the {\it Dyson's density } 
\begin{equation}
\nu _D\left( \varepsilon \right) =\frac 2{\pi ^2}{\cal P}\int_0^{D_N}\frac{%
tdt}{t^2-\varepsilon ^2}\frac{dV}{dt}\sqrt{\frac{1-\varepsilon ^2/D_N^2}{%
1-t^2/D_N^2}}  \label{eq.21}
\end{equation}
{\it defined on the whole real axis} {\bf R}: 
\begin{equation}
A_N\left( \varepsilon \right) =\frac{\pi \nu _D\left( \varepsilon \right) }{%
\sqrt{1-\varepsilon ^2/D_N^2}}\text{.}  \label{eq.22}
\end{equation}
Here $\varepsilon $ takes {\it arbitrary} value that can lie both inside and
outside of an eigenvalue support. The spectrum endpoint $D_N=2c_N$ is the
positive root of the integral equation 
\begin{equation}
N=\frac 2\pi \int_0^{D_N}\frac{dV}{dt}\frac{tdt}{\sqrt{D_N^2-t^2}}
\label{eq.23}
\end{equation}
following from normalization of Dyson's density.

Combining Eqs. (\ref{eq.12}), (\ref{eq.17}), (\ref{eq.22}), and using
asymptotic identity 
\begin{equation}
B_N=\frac \varepsilon {D_N}A_N-\frac{dV}{d\varepsilon },  \label{eq.24}
\end{equation}
which is a consequence of Eqs. (\ref{eq.15}) and (\ref{eq.16}), it is a
straightforward step to reach the following remarkable asymptotic
differential equation: 
\begin{equation}
\psi _N^{\prime \prime }-\left[ \frac d{d\varepsilon }\log \left( \frac{\pi
\nu _D\left( \varepsilon \right) }{\sqrt{1-\varepsilon ^2/D_N^2}}\right)
\right] \psi _N^{\prime }+\pi ^2\nu _D^2\left( \varepsilon \right) \psi _N=0%
\text{,}  \label{eq.25}
\end{equation}
which together with relationship 
\begin{equation}
\psi _N^{\prime }=\frac{\pi \nu _D\left( \varepsilon \right) }{\sqrt{%
1-\varepsilon ^2/D_N^2}}\left( \psi _{N-1}-\frac \varepsilon {D_N}\psi
_N\right)   \label{eq.26}
\end{equation}
provide a general basis for the study of eigenvalue correlations in {\it %
arbitrary spectral range }\cite{remark-2}.

An interesting property of these equations is that they do not contain
confinement potential explicitly, but only involve the {\it Dyson's density} 
$\nu _D$ and spectrum endpoint $D_N$. Moreover, it turns out that the
knowledge of Dyson's density (that coincides with {\it real} density of
states only in the spectrum bulk) is sufficient to determine the {\it genuine%
} density of states, as well as the $n-$point correlation function, {\it %
everywhere}. We also note that Eq. (\ref{eq.25}) can be derived in a
different way for monotonous confinement potentials increasing at least as
fast as $\left| \varepsilon \right| $ at infinity. This suggests that
differential equation Eq. (\ref{eq.25}) should hold generally and not only
for confinement potentials having polynomial form Eq. (\ref{eq.11}).

{\bf B. }Up to this point our derivation was quite general without any
respect to the soft edge of eigenvalue support. We now focus our attention
on the eigenvalue correlations near the soft edge $\varepsilon =D_N$. It is
known \cite{Bowick} that tuning coefficients $d_k$ which enter $V$ one can
reach a situation when the bulk (Dyson's) density of states will possess a
singularity of the type: 
\begin{equation}
\nu _D\left( \varepsilon \right) =\left[ 1-\frac{\varepsilon ^2}{D_N^2}%
\right] ^{m+1/2}{\cal R}_N\left( \frac \varepsilon {D_N}\right) 
\label{eq.27}
\end{equation}
with $m=0,2,4,$ etc., and ${\cal R}_N$ being a well-behaved function with $%
{\cal R}_N\left( 1\right) \neq 0$. [Odd indices $m$ are inconsistent with
our choice for leading coefficient $d_p$, entering confinement potential $%
V\left( \varepsilon \right) $, be positive in order to keep a convergence of
integral for partition function ${\cal Z}_N$ in Eq. (\ref{eq.01})]. We
intend to demonstrate that as long as multicriticality of order $m$ is
reached, the eigenvalue correlations in the vicinity of the soft edge become
universal, and are independent of the particular potential chosen. The order 
$m$ of the multicriticality is the only parameter which governs spectral
correlations in the soft-edge scaling limit.

Let us move the spectrum origin to its endpoint $D_N$, making replacement 
\begin{equation}
\varepsilon _s=D_N\left[ 1+s\cdot \frac 12\left( \frac 2{\pi D_N{\cal R}%
_N\left( 1\right) }\right) ^{1/\nu ^{*}}\right] ,  \label{eq.28}
\end{equation}
that defines the $m-$th {\it soft-edge scaling limit} provided $s\ll \left(
D_N{\cal R}_N\left( 1\right) \right) ^{1/\nu ^{*}}\propto N^{1/\nu ^{*}}$.
It is straightforward to show from Eqs. (\ref{eq.25}) and (\ref{eq.26}) that
the function $\widehat{\psi }_N\left( s\right) =\psi _N\left( \varepsilon
_s-D_N\right) $ obeys differential equation 
\begin{equation}
\widehat{\psi }_N^{\prime \prime }\left( s\right) -\frac{\left( \nu ^{*}-%
\frac 32\right) }s\widehat{\psi }_N^{\prime }\left( s\right) -s^{2\left( \nu
^{*}-1\right) }\widehat{\psi }_N\left( s\right) =0,  \label{eq.29}
\end{equation}
and that the following relation takes place: 
\begin{eqnarray}
\widehat{\psi }_{N-1}\left( s\right)  &=&\widehat{\psi }_N\left( s\right) 
\label{eq.30} \\
&&+\left( -1\right) ^{\nu ^{*}-\frac 32}\left( \frac 2{\pi D_N{\cal R}%
_N\left( 1\right) }\right) ^{\frac 1{2\nu ^{*}}}s^{\frac 32-\nu ^{*}}%
\widehat{\psi }_N^{\prime }\left( s\right) .  \nonumber
\end{eqnarray}

Solution to Eq. (\ref{eq.29}) which decreases at $s\rightarrow +\infty $
(that is at far tails of the density of states) is given (up to an arbitrary
factor $\lambda _N$) by the function $%
\mathop{\rm G}
\left( s|\nu ^{*}\right) $, Eq. (\ref{eq.06}). The factor $\lambda _N$ can
be found by fitting \cite{Airy} the density of states $K_N\left( \varepsilon
_s,\varepsilon _s\right) $, Eq. (\ref{eq.08}), to the bulk density of
states, Eq. (\ref{eq.27}), near the soft edge provided $1\ll s\ll N^{1/\nu
^{*}}$. Then, making use of Eqs. (\ref{eq.06}), (\ref{eq.08}) and (\ref
{eq.30}) we easily obtain that in the $m-$th soft-edge scaling limit, Eq. (%
\ref{eq.28}), the two-point kernel 
\begin{equation}
K_{\text{soft}}^{\left( m\right) }\left( s,s^{\prime }\right)
=\lim_{N\rightarrow \infty }K\left( \varepsilon _s,\varepsilon _{s^{\prime
}}\right) \frac{d\varepsilon _s}{ds}  \label{eq.31}
\end{equation}
is determined by Eq. (\ref{eq.05}). In particular case of $m=0$, that is
inherent in random-matrix ensembles with monotonous confinement potential,
the function $%
\mathop{\rm G}
$ coincides with Airy function, $%
\mathop{\rm G}
\left( s|\frac 32\right) =%
\mathop{\rm Ai}
\left( s\right) $, and the Airy correlations, Eq. (\ref{eq.04}), are
recovered.

It follows from Eqs. (\ref{eq.05}) and (\ref{eq.29}) that density of states
in the same scaling limit 
\begin{equation}
\nu _{\text{soft}}^{\left( m\right) }\left( s\right) =\left( \frac d{ds}%
\mathop{\rm G}
\left( s|\nu ^{*}\right) \right) ^2s^{\frac 32-\nu ^{*}}-\left[ 
\mathop{\rm G}
\left( s|\nu ^{*}\right) \right] _{}^2s^{\nu ^{*}-\frac 12}  \label{eq.32}
\end{equation}
is also universal.

The large-$\left| s\right| $ behavior of $\nu _{\text{soft}}^{\left(
m\right) }$ can be deduced from the known asymptotic expansions of the
Bessel functions: 
\begin{equation}
\nu _{\text{soft}}^{\left( m\right) }\left( s\right) =\left\{ 
\begin{array}{ll}
\frac{\left| s\right| ^{\nu ^{*}-1}}\pi +\frac{\left( -1\right) ^{\nu ^{*}-%
\frac 12}}{4\pi \left| s\right| }\cos \left( \frac{2\left| s\right| ^{\nu
^{*}}}{\nu ^{*}}\right) , & s\rightarrow -\infty , \\ 
\frac{\exp \left( -\frac{2s^{\nu ^{*}}}{\nu ^{*}}\right) }{4\pi s}\frac{\cos
^2\left( \frac \pi {4\nu ^{*}}\right) }{\sin \left( \frac \pi {4\nu ^{*}}%
\right) +\left( -1\right) ^{\nu ^{*}-\frac 32}}, & s\rightarrow +\infty .
\end{array}
\right.   \label{eq.33}
\end{equation}
Note that the leading order behavior as $s\rightarrow -\infty $ is
consistent with the $\left| s\right| ^{\nu ^{*}-1}$ singularity of the bulk
density of states, Eq. (\ref{eq.27}).

To conclude, in this paper we presented a general formalism for a treatment
of eigenlevel correlations in spectra of $U\left( N\right) $ invariant
ensembles of large random matrices with strong level confinement. An
important ingredient of our analysis is the second order differential
equation which connects the Dyson's density with a fictitious
``wavefunction'' $\psi _N$ which is needed for calculations of eigenvalue
correlations within the framework of orthogonal polynomial technique. This
consideration is relevant to arbitrary energy range. We have applied this
formalism to examine the eigenlevel correlations near the endpoint of single
spectrum support. It has been shown that in the soft-edge scaling limit
there are novel universal eigenlevel correlations which only depend on the
even multicritical index of a matrix model. In a particular case $m=0$,
corresponding to monotonous confinement potentials, universality of the Airy
correlations is recovered.

One of the authors (E. K.) acknowledges the support of the Levy Eshkol
Fellowship from the Ministry of Science of Israel.

\widetext


\begin{references}
\bibitem{remark-1}  Namely, $V\left( \varepsilon \right) $ should increase
at least as fast as $\left| \varepsilon \right| $ at infinity, see Refs. 
\cite{Weidenmueller,Kanzieper}.
\end{references}
\end{document}